# Hysteresis and the dynamic phase transition in thin ferromagnetic films

Hyunbum Jang[1] and Malcolm J Grimson[2]
[1]Department of Chemical Engineering, North Carolina State University, Raleigh, USA
[2]Department of Physics, University of Auckland, Auckland, New Zealand

Hysteresis and the non-equilibrium dynamic phase transition in thin magnetic films subject to an oscillatory external field have been studied by Monte Carlo simulation. The model under investigation is a classical Heisenberg spin system with a bilinear exchange anisotropy $\Lambda$ in a planar thin film geometry with competing surface fields. The film exhibits a non-equilibrium phase transition between dynamically ordered and dynamically disordered phases characterized by a critical temperature $T_{cd}$, whose location of is determined by the amplitude $H_0$ and frequency $\omega$ of the applied oscillatory field. In the presence of competing surface fields the critical temperature of the ferromagnetic-paramagnetic transition for the film is suppressed from the bulk system value, $T_c$, to the interface localization-delocalization temperature $T_{ci}$. The simulations show that in general $T_{cd} < T_{ci}$ for the model film. The profile of the time-dependent layer magnetization across the film shows that the dynamically ordered and dynamically disordered phases coexist within the film for $T < T_{cd}$. In the presence of competing surface fields, the dynamically ordered phase is localized at one surface of the film.

## I. Introduction

When a ferromagnet is subject to a time dependent oscillatory external field $H(t)$, the system cannot typically respond instantaneously. Thus the time dependent magnetization of the system lags behind the driving field and hysteresis results. The area of the hysteresis loop $A$ is equal to the energy dissipated per period of the applied oscillatory field and its dependence on the frequency and amplitude of the applied field has been extensively studied. An ultra-thin ferromagnetic film with a uniaxial anisotropy that is driven by a oscillatory external field $H(t) = H_0 \sin(\omega t)$, where $H_0$ is the amplitude and $\omega$ is the angular frequency of the applied field, will switch between two stable states of positive and negative magnetization that are degenerate in the absence of the applied field ($H_0 = 0$). Experimental studies [1-4] have observed power law scaling of the hysteresis loop area $A$ with $A \sim H_0^\alpha \omega^\beta$ which was generally consistent with mean-field theory [5,6] and early Monte Carlo studies of the kinetic Ising model [7-11]. However, there was much disagreement in the reported values for the exponents $\alpha$ and $\beta$.

Subsequent extensive Monte Carlo simulations of the kinetic Ising model [12-14] have shown that the hysteresis loop area exhibits an extremely slow approach to an asymptotic, logarithmic dependence on the product of the amplitude and the field frequency. This may explain the inconsistent exponent estimates reported in attempts to fit experimental and numerical data for the low frequency behavior of the hysteresis loop area to a power law. At higher frequencies a dynamic phase transition is observed in which the period averaged magnetization $Q$ passes from a dynamically disordered state with $Q = 0$ to a dynamically ordered state with $Q > 0$. This dynamic phase transition can be intuitively understood as the competition between two time scales: the period of the applied oscillatory field and the response time of the magnetization. When the field oscillates at sufficiently low frequency the magnetization essentially follows the field, switching the system between its two zero-field stable states with the same period as the applied field, provided that the amplitude of the external force is sufficiently large. At higher frequencies, the system is unable to relax quickly enough even to follow the sign (phase) of the external field and settles down into a symmetry–breaking oscillation about one or other of its zero-field stable states. The location of the transition is a function of the temperature, field amplitude and frequency. A finite-size scaling analysis of large-scale Monte Carlo simulations of the kinetic Ising model in an oscillatory field has shown that the dynamic phase transition is in the same universality class as the equilibrium Ising model [15]. A result confirmed in a recent study of a time-dependent Ginzburg-Landau model in an oscillatory field [16].

The study of thin film ferromagnetism is of intense significance. Not only for its applications in magnetic recording media, a key component of today's information technology industry, but also for the fundamental physics it reveals. Finite size effects in thin films arising from both confinement and surface modification give rise to a variety of novel equilibrium phase behaviors that are not observed in the bulk materials. In this context, the interface localization-delocalization transition in thin ferromagnetic films with competing surfaces has been the subject of much recent investigation. The competing surface forces are surface anisotropies in the direction perpendicular to the plane of the film that favor a positive magnetization at one surface and a negative magnetization at the other surface. Binder *et al*. [17-20] have made an extensive study of the thin ferromagnetic Ising film with competing surface forces and shown that the properties of the interface localization-delocalization transition are distinct from both the bulk ferromagnetic-paramagnetic phase transition and the wetting transition in semi-infinite systems. Complementary studies on thin ferromagnetic Heisenberg films with competing surface forces [21,22] have shown that the presence of an interface localization-delocalization transition is not restricted to discrete state models, such as the Ising model. But it is also found in magnetic systems where the spins are continuously orientable, albeit with some degree of uniaxial anisotropy.

While the kinetic Ising model can provide a good model of uniaxial ferromagnets in which magnetization reversal proceeds by nucleation and domain wall motion. It cannot account for magnetic relaxation processes such as the coherent rotation of spins. This requires a spin model with continuous degrees of freedom such as the classical Heisenberg model in which the magnetic spins can rotate through all possible orientations. However, studies of the magnetic phase behavior of the Heisenberg model are more complicated than for the corresponding Ising model. With only isotropic interactions between nearest neighbor spins, ferromagnetic order is only found at zero temperature in the absence an external field. However the inclusion of a uniaxial anisotropy in the Hamiltonian can significantly modify the properties of the Heisenberg spin system. A uniaxial anisotropy favors the alignment of spins along an easy axis, conventionally denoted as the $z$-axis, which can be regarded as the $c$-axis in hexagonal, tetragonal and rhombohedral crystals. For sufficiently large values of the uniaxial anisotropy, Ising-like phase behavior is recovered [21,22].

This paper investigates hysteresis and the dynamic phase behavior of thin ferromagnetic films within the anisotropic Heisenberg model. The inclusion of competing surface fields allows the magnetization distribution within the film to be controlled and its interplay with driving force provided by the applied oscillatory field studied. In the following section a full description of the model is given together with the details of the Monte Carlo simulation method. In section III the dynamic phase behavior of the thin ferromagnetic film with free surfaces is presented along with the results for the corresponding bulk system. Modifications to the dynamic phase behavior resulting from the addition of competing surface fields are given in section IV. The structure of the magnetization within the film is detailed in section V and the role of the frequency of the applied oscillatory fields is the topic of section VI. The paper closes with a conclusion.

## II. The model

The Hamiltonian for the classical Heisenberg model with a bilinear exchange anisotropy $\Lambda$ can be written as [22]

$$H_0 = -J \sum_{\langle i,j \rangle} \left( (1-\Lambda)\left(S_i^x S_j^x + S_i^y S_j^y\right) + S_i^z S_j^z \right). \tag{1}$$

where $\mathbf{S}_i = (S_i^x, S_i^y, S_i^z)$ is a unit vector representing the $i$th spin and the notation $\langle i,j \rangle$ means that the sum is restricted to nearest-neighbor pairs of spins. $J$ is a coupling constant characterizing the magnitude of the exchange interaction and for ferromagnets $J > 0$. Following Binder and Landau [23], $\Lambda$ determines the strength of the bilinear exchange anisotropy and is only applied to the $x$ and $y$ components of the spin. In the isotropic limit, $\Lambda = 0$, the model reduces to the familiar classical Heisenberg model of magnetism, while for $\Lambda = 1$, the Hamiltonian becomes Ising-like.

The system under consideration here is a three dimensional thin planar film of finite thickness $D$ subject to applied surface fields and an oscillatory external field with Hamiltonian

$$H = H_0 - \sum_{i \in \text{surface } 1} \mathbf{H}_1 \cdot \mathbf{S}_i - \sum_{i \in \text{surface } D} \mathbf{H}_D \cdot \mathbf{S}_i - H(t) \sum_i S_i^z, \tag{2}$$

where $\mathbf{H}_1$ and $\mathbf{H}_D$ are the applied surface fields. The time dependent oscillatory external field $H(t)$ is taken to have a sinusoidal form with

$$H(t) = H_0 \sin(\omega t), \tag{3}$$

where $H_0$ is the amplitude and $\omega$ is the angular frequency of the oscillatory field.

We consider a simple cubic lattice of size $L \times L \times D$, in units of the lattice spacing, and apply periodic boundary condition in the $x$ and $y$ directions. Free boundary conditions are applied in the $z$ direction that is of finite thickness $D$. The system is subject to competing applied surface fields in layers $n = 1$ and $n = D$ of the film with

$$\mathbf{H}_1 = h\hat{\mathbf{z}}\delta_{i1}, \tag{4}$$
$$\mathbf{H}_D = -h\hat{\mathbf{z}}\delta_{iD}, \tag{5}$$

giving a Hamiltonian

$$\mathcal{H} = H_0 - h\left(\sum_{i \in \text{surface 1}} S_i^z - \sum_{i \in \text{surface } D} S_i^z\right) - H(t)\sum_i S_i^z. \tag{6}$$

A film thickness $D = 12$ was used throughout. The value of $D = 12$ corresponds to the crossover regime between wall and bulk dominated behavior for thin Ising films [18]. In thinner films it is difficult to distinguish between "interface" and "bulk" phases in the film, since all layers of the film feel the effect of the competing surface fields rather strongly. While for thicker films the surfaces of the film only interact close to the bulk critical point. Results are reported here for lattices of size $L = 32$, but no significant differences were found for lattices with $L = 64$ and $L = 128$ at non-critical values of $H_0$, $\omega$ and $T$. The Metropolis algorithm [24] was used in the Monte Carlo simulations with trial configurations generated from Barker-Watts [25] spin rotations. In the simulations trial spin rotations were performed sequentially through the lattice in a checkerboard fashion. One full scan of the entire lattice comprises one Monte Carlo step per spin (MCSS), the unit of time in the simulations. The period of the sinusoidal external field is given by the product $FSR \times N$, where $FSR$ is the field sweep rate [26] and $N$ is the number of MCSS. The applied oscillatory field $H(t)$ being updated after every MCSS according to Eq. (3). The majority of the simulations were performed for a value of $FSR = 1$ with $N = 240$. For lower frequencies of the applied oscillatory field, larger values for $FSR$ were used up to a value of $FSR = 1000$. In all the simulations a random initial spin configuration was used.

No significant changes to the dynamical properties reported in this paper were found when a random spin update scheme replaced the checkerboard sequential updating used in these simulations of the classical Heisenberg spin system. Although it should be noted that Monte Carlo studies of the very fine detail in the dynamics for the kinetic Ising model have revealed significant differences between random and sequential spin updating schemes [27].

The time-dependent magnetic order of the film is characterized by $z$ component of the magnetization for the film

$$M_z(t) = \frac{1}{D} \sum_{n=1}^{D} M_n^z(t) \qquad (7)$$

and the time-dependent z component of the magnetization for the *n*th layer of the film

$$M_n^z(t) = \frac{1}{L^2} \sum S_i^z(t) \qquad (8)$$

were calculated during the simulations. The order parameter *Q* for the dynamic phase transition [11] is the period-averaged magnetization over a complete cycle of the sinusoidal field defined by

$$Q = \frac{\omega}{2\pi} \oint M_z(t)\, dt, \qquad (11)$$

The hysteresis loop area *A* is defined by

$$A = -\oint M_z\, dH. \qquad (12)$$

### III. Free film

In this paper we have focused on a system with a bilinear exchange anisotropy of $\Lambda = 0.1$. For this weak exchange anisotropy $\Lambda$ the system is intermediate in character between the limiting Ising-like ($\Lambda = 1$) and Heisenberg ($\Lambda = 0$) models. In the absence of an applied field the bulk system displays a second order ferromagnetic – paramagnetic phase transition at $T_c^* = k_B T_c/J = 1.53$. For the thin film geometry considered here with film thickness $D = 12$ and in the absence of any applied field the critical temperature characterizing the ferromagnetic-paramagnetic phase transition is reduced with $T_c^* = 1.51$ for the free film.

When subject to an applied oscillatory field $H(t)$ the magnetization of the film becomes time dependent. The dynamic response of the film is characterized by the period-averaged magnetization, *Q*, and the hysteresis loop area, *A*. Fig. 1 shows $\langle Q \rangle$ and $\langle A \rangle$ as a function of the field amplitude $H_0$ for an applied oscillatory field with angular frequency $\omega = 2\pi / 240$. $\langle Q \rangle$ and $\langle A \rangle$ are averaged values over a sequence of full cycles with initial transients discarded. The number of cycles in the average was adjusted to ensure the statistical average was much smaller than the symbol in the figure. The figure presents the results of simulations at reduced temperature of $T^* = k_B T/J = 1.0$. For comparative purposes the figure also shows the corresponding result for the bulk system subject to the same applied oscillatory field. These results were obtain from a simulation of a $32 \times 32 \times 32$ simple cubic lattice with periodic boundary conditions. The qualitative form of $\langle Q \rangle$ and $\langle A \rangle$ as a function of $H_0$ in Fig. 1 is the same for both the bulk system and the free film. For small $H_0$, $\langle Q \rangle$ is a constant and $\langle A \rangle = 0$. Since $T^* < T_c^*$, the system is ferromagnetic. For sufficiently small $H_0$, the applied field is too weak to produce any significant reorientation of the spins in the ferromagnetically ordered sample. Hence the film magnetization is essentially time independent. As a consequence $\langle A \rangle = 0$ and the non-zero magnetization of the ferromagnetically ordered system ensures $\langle Q \rangle$ has a non-zero value for small $H_0$.

As $H_0$ increases the driving force of the oscillatory field starts to dominate the competing ferromagnetic ordering arising from the spin-spin interactions. The alignment of spins then tends to follow the oscillatory field and gives rise to temporal oscillations in the film magnetization with an angular frequency consistent with the applied field. As a result $\langle Q \rangle$ decreases, while $\langle A \rangle$ increases, with increasing $H_0$. For sufficiently large $H_0$ the applied field becomes so dominant that $\langle Q \rangle$ vanishes as a result of the symmetric variation of the time-dependent magnetization. The dynamic phase transition is characterized by the order parameter $Q$ that vanishes at a non-zero value of $H_0$ with increasing $H_0$. Note that the location of the dynamic phase transition in the free film is at a slightly lower value of $H_0$. Fig. 1 showing that $\langle Q \rangle$ vanishes at values of $H_0 = 0.76$ in the free film and $H_0 = 0.80$ in the bulk system. For values of $H_0$ greater than these critical values, there is a smooth monotonic increase in $\langle A \rangle$ with increasing $H_0$ which has a power law form with $\langle A \rangle \sim H_0^\alpha$. The results in Fig.1 correspond to values of $\alpha = 0.75$ in the bulk system and $\alpha = 0.73$ in the free film. Such values for $\alpha$ are slightly higher than those found in comparable Ising model studies [11]. Although estimates for power law scaling exponents extracted from fits to data over a restricted amplitude range must be treated with caution [14].

The temperature dependence of the period averaged magnetization $Q$ and the hysteresis loop area $A$ is shown in Fig. 2 for the free film and the bulk system at an amplitude of the applied oscillatory field, $H_0 = 1.0$. Fig. 2 shows that $\langle Q \rangle \neq 0$ at low $T^*$ and the system is in a dynamically ordered phase. While a dynamically disordered state is found at high $T^*$ with $\langle Q \rangle = 0$. The critical temperature characterizing the dynamic phase transition, $T_{cd}^*$, is lower in the free film than in the bulk system with $T_{cd}^* = 0.81$ for the free film and $T_{cd}^* = 0.86$ for the bulk system. Note that $T_{cd}^* \ll T_c^*$ for both the free film and the bulk system. Fig. 2 further shows that the dependence of $\langle A \rangle$ on $T^*$ is qualitatively different from the dependence of $\langle A \rangle$ on $H_0$ seen in Fig. 1. For fixed $H_0$, Fig. 2 shows that $\langle A \rangle$ as a function of $T^*$ possesses a broad, but clear, maximum located temperatures just above $T_{cd}^*$. These results are consistent with studies of the two dimensional kinetic Ising model [11, 14] and provide additional evidence for the existence of a dynamic phase transition at $T_{cd}^*$.

Additional studies of smaller systems with $L = 16$ returned results consistent with those presented here for a lattice with $L = 32$. But for very small systems with $L = 6$, there is no significant lag of the spins behind the applied oscillatory field and no dynamic phase transition is observed [12].

## IV. Competing surface fields

For the thin film geometry considered in this paper with a film thickness $D = 12$ and competing surface fields with $h = 0.55$, the system exhibits an interface localization - delocalization transition at a critical temperature of $T_{ci}^* = 1.12$ in the absence of an applied field. This is well below the critical temperature of ferromagnetic – paramagnetic phase transition for the bulk system, where $T_c^* = k_B T_c/J = 1.53$. Thus for the system under consideration here, the order – disorder and interface localization - delocalization phase transitions are quite distinct.

The magnetization of the film becomes time dependent when subject to an applied oscillatory field. The dynamic response of the film with competing surface fields is given in Fig. 3. This shows $\langle Q \rangle$ and $\langle A \rangle$ as a function of the field amplitude $H_0$ for an applied oscillatory field with angular frequency $\omega = 2\pi / 240$. The figure presents the results of simulations at reduced temperatures of $T^* = k_B T/J = 0.6, 1.0,$ and $1.2$. In equilibrium with $H_0 = 0$, the corresponding bulk systems are ferromagnetic at all three temperatures, while the film displays a localized interface in the magnetization profile at the lowest temperature and a delocalized interface at the highest temperature. The qualitative form of $\langle Q \rangle$ and $\langle A \rangle$ as a function of $H_0$ in Fig. 3 is the same for all three temperatures *i.e.* for thin ferromagnetic films above and below the interface localization - delocalization transition. In the limit $H_0 \to 0$, the net magnetization of the film with competing surface fields is zero and hence $\langle Q \rangle = 0$ for $H_0 \to 0$. For small $H_0$ there is an, initially linear, increase in $\langle Q \rangle$ with $H_0$ up to a maximum in $\langle Q \rangle$. As $H_0$ increases further, $\langle Q \rangle$ decreases to zero.

The dynamic phase transition is characterized by the order parameter $Q$ that vanishes at a non-zero value of $H_0$ with increasing $H_0$. For increasing $T^*$ the location of the dynamic phase transition shifts to lower values of $H_0$. Fig. 3 showing that at temperatures of $T^* = 0.6, 1.0$ and $1.2$, $\langle Q \rangle$ vanishes at values of $H_0 = 1.3, 0.75$ and $0.5$ respectively. In addition the peak in $\langle Q \rangle$ decreases in magnitude while its location also shifts to lower $H_0$ with increasing $T^*$. All direct consequences of the greater thermal disorder in the spin system at higher $T^*$ requiring a smaller amplitude of the applied field to dominate the ferromagnetic order of the sample and drive the dynamic reorientation of the spins. For small $H_0$, Fig.3 shows $\langle A \rangle = 0$ and a smooth monotonic increase in $\langle A \rangle$ with increasing $H_0$ coinciding with the decrease to zero of $\langle Q \rangle$. For values of $H_0$ above the dynamic phase transition, the results of Fig.3 are consistent with power law scaling of the form $\langle A \rangle \sim H_0^\alpha$. The exponent $\alpha$ shows no significant dependence on with $\alpha = 0.74, 0.73$ and $0.72$ for $T^* = 0.6, 1.0$ and $1.2$ respectively. Such values for $\alpha$ are consistent with those obtained for the free film and bulk system reported in section III, although estimates for the power law scaling exponent $\alpha$ determined from a restricted amplitude range must be treated with caution [14].

The temperature dependence of the period averaged magnetization $Q$ and the hysteresis loop area $A$ in the film with competing surface fields is shown in Fig. 4 for three amplitudes of the applied oscillatory field: $H_0 = 0.3, 0.55$ and $1.0$. So the figure contains information on the dynamic phase transition in the ferromagnetic films for applied oscillatory fields whose magnitude is below, equal to and above that of the surface field $h = 0.55$. For all values of $H_0$, Fig. 4 shows that $\langle Q \rangle \neq 0$ at low $T^*$ and the system is in a dynamically ordered phase. While a dynamically disordered state is found at high $T^*$ with $\langle Q \rangle = 0$. However the critical temperature characterizing the dynamic phase transition, $T_{cd}^*$, is $H_0$ dependent. Fig. 4 shows $T_{cd}^* = 1.43, 1.18$ and $0.80$ for the oscillatory field amplitudes of $H_0 = 0.3, 0.55$ and $1.0$ respectively. Thus the critical temperature for the dynamic phase transition of the film decreases with increasing $H_0$. Fig. 4 further shows that the hysteresis loop area $\langle A \rangle$ as a function of $T^*$ possesses a broad, but clear, maximum located temperatures just above $T_{cd}^*$ as previously seen for the bulk system and the free film with no surface fields.

As the magnitude of the surface field strength $h$ is increased, the critical temperature of the interface localization – delocalization transition $T_{ci}^*$ decreases. However our simulations show that $T_{cd}^*$ also decreases as $h$ increases and that $T_{cd}^* < T_{ci}^*$ for all $h$. Indeed, the qualitative form of the hysteresis curves is essentially independent of the magnitude of the competing surface fields.

A greater insight into the nature of the dynamic phase transition seen in Figs. 3 and 4 follows from the time dependence of the $z$ component of the magnetization, $M_z(t)$, in the applied oscillatory field $H(t)$. Fig. 5 shows $M_z(t)$ for the film over the initial cycles of the applied oscillatory field at a temperature $T^* = 1.0$ for an applied oscillatory field of angular frequency $\omega = 2\pi / 240$ with amplitudes $H_0 = 0.3, 0.7, 1.0, 2.0$, and $3.0$. The figure shows that memory of the initial state in the simulation is short, the steady state being rapidly obtained within a few cycles of the applied oscillatory field for all $H_0$. From Fig. 3 it can be seen that $H_0 = 0.3$ and $T^* = 1.0$ corresponds to a dynamically ordered state with a non-zero value of $Q$. Fig. 5 shows after initial transients $M_z$ for $H_0 = 0.3$ and $T^* = 1.0$ has an oscillatory form with the same angular frequency as the applied oscillatory field, but lags behind the field by approximately $\pi/2$. The dynamic ordering of the state is evidenced by the non-zero mean value for $M_z(t)$. The hysteresis curve is obtained by plotting $M_z(t)$ in the $M_z$–$H$ plane and for $H_0 = 0.3$ and $T^* = 1.0$ the hysteresis curve is the asymmetric loop shown in Fig. 6(a). The asymmetric loop can be located in either the positive or negative $M_z$ half-plane depending on the initial direction of the applied oscillatory field. For the increased field amplitude $H_0 = 0.7$, the time-dependent magnetization still smoothly oscillates with the applied field but with a slight decrease in the phase lag of $M_z$ behind $H$. Most notably though the mean value of $M_z(t)$ is markedly reduced. The hysteresis curve for $H_0 = 0.7$ and $T^* = 1.0$ from the data in Fig. 5 is given in Fig. 6(b) and is close to symmetric about the $M_z$ and $H$ axes. Thus the period averaged magnetization $\langle Q \rangle$ is very small, while the hysteresis loop area is much larger than for $H_0 = 0.3$. From Fig. 3 it can be seen that the $H_0 = 0.7$ and $T^* = 1.0$ state is located in the vicinity of the phase transition between dynamically ordered and dynamically disordered states.

When the applied oscillatory field amplitude increases further to $H_0 = 2.0$ and $3.0$, the time delayed dynamic response of $M_z(t)$ to the oscillatory field at $T^* = 1.0$ shown in Fig. 5 corresponds to the hysteresis loops as in Fig 6(c) and (d). These systems are characterized by a zero value for the period averaged magnetization and large values for the hysteresis loop area that increase with $H_0$. The onset of saturation in the peaks and troughs of $M_z(t)$ results in the more angular shape for the hysteresis loop that becomes more marked with increasing $H_0$. Note that the lag in the response of $M_z$ to the applied field $H$ reduces with increasing $H_0$.

Fig. 7 show simulation results for the time variation of $z$ component magnetization $M_z(t)$ with an applied field of amplitude $H_0 = 1.0$ and angular frequency $\omega = 2\pi / 240$ for temperatures $0.6 \leq T^* \leq 1.4$. The hysteresis curves corresponding to the results for $M_z(t)$ in Fig. 7 at $T^* = 0.6, 0.8, 1.0$ and $1.4$ are shown in Fig. 8. From Fig. 4 for $H_0 = 1.0$, the critical temperature for the dynamic phase transition is given by $T_{cd}^*(H_0=1.0) = 0.8$. For $T < T_{cd}$, $M_z(t)$ lags $H(t)$ by approximately $\pi/2$ and the dynamically ordered phase of the film is characterized by a non-zero mean value for $M_z(t)$. The asymmetric loop in the hysteresis curve can be located in either the positive or negative $M_z$ half-plane depending on the initial direction of the applied oscillatory field. See Fig. 8(a) for $T^* = 0.6$. Fig. 8 (b) shows that the hysteresis loop becomes essentially symmetric about the $M_z$ and $H$ axes around $T_{cd}$. This gives rise to the vanishing of $Q$ and indicates the onset of a dynamically disordered phase. For $T > T_{cd}$, the phase lag of $M_z(t)$ behind $H(t)$ decreases with increasing $T$. This is a result of thermal disorder reducing the ferromagnetic ordering tendencies of the spins which become more able to respond to the applied field. Furthermore, another consequence of enhanced thermal disorder with increasing $T$ is that the magnitude of $M_z(t)$ in the cycle of $H(t)$ decreases with increasing $T$. This reduces the magnitude of hysteresis loop area for $T > T_{cd}$. As a result the hysteresis loop area has maximum value at temperatures just above $T_{cd}^*$.

## V. Dynamic response of the layer magnetization within the film

Further information on the nature of the dynamic phase behavior of the film seen in section IV is contained in the layer magnetization across the film. Fig. 9 shows the time-dependent layer magnetization across the film, $M_n^z(t)$, over three consecutive cycles of the applied oscillatory field for $(H_0, T^*) = (1.0, 1.0), (1.0, 0.6)$ and $(3.0, 1.0)$. From the results of section IV it can be seen that the states $(H_0, T^*) = (1.0, 1.0)$ and $(3.0, 1.0)$ correspond dynamically disordered states of the film, while for $(H_0, T^*) = (1.0, 0.6)$ the film is in a dynamically ordered state. A qualitative difference between the results for $M_n^z(t)$ in the dynamically ordered phase (Fig. 9(b)) and those in the disordered phase (Fig. 9(a) and (c)) is immediately apparent. For the dynamically disordered states of the film, $M_n^z(t)$ has the same qualitative form for all $n$. The time dependent layer magnetization is essentially uniform across the film and at any instant of time no interface between regions of negative and positive magnetization of the film can be observed. The "uniform" time-delayed response of the layer magnetization across the film to the applied oscillatory field leads to a symmetric hysteresis loop for the film and a zero value for the dynamic order parameter of the film $Q$.

This is in marked contrast to the behavior of the dynamically ordered system shown in Fig. 9(b). Here $M_n^z(t)$ at the near surface of the film, layer $n = 1$, oscillates between positive and negative values in a time-delayed response to the applied oscillatory field. The time-averaged magnetization in layer 1 is close to zero. Indeed the behavior in layer $n = 1$ is reminiscent of that seen in all layers of the film in a dynamically disordered state. But $M_n^z(t)$ at the far surface of the film, layer $n = 12$, is markedly different. It shows a large non-zero value for the time-averaged magnetization of the layer with only a very weak oscillatory response to the applied oscillatory field. So that the behavior of layer n = 12 is akin to that observed in the dynamically ordered state. Thus for an applied oscillatory field with $H_0 = 1.0$, at a temperature $T^* = 0.6$, dynamically ordered and dynamically disordered states coexist within the film. From Fig. 4 for $H_0 = 1.0$, it can be seen that the temperature $T^* = 0.6$ is below the dynamic critical temperature $T_{cd}^* = 0.8$. The non-zero value of $Q$ for the film in this state arises from the contribution of the dynamically ordered layers on the far side of the film, $n > 4$. For $T^* > T_{cd}^*$ all layers of the film correspond to dynamically disordered states. But when $T^*$ is below $T_{cd}^*$ dynamically ordered layers form in the film. The dynamically ordered layers are located near the far surface of the film since $T^* < T_{cd}^* < T_{ci}^*$ and the underlying equilibrium state of the film is an interface localized state. In this case one with a net positive magnetization of the film. Note however that the net magnetization of the film in an interface-localized state can be either positive or negative depending on the initial spin configuration used in the simulation and the phase constant of the applied oscillatory field.

## VI. Field sweep rate

Recent experimental work [1-4] and theoretical studies of the kinetic Ising model [11-14] have shown that the form of the hysteresis loop strongly depends on the frequency of the applied oscillatory field. To conclude this work, the dependence of the hysteresis curve on the frequency of the applied oscillatory field is investigated. Fig. 8(c) shows the hysteresis loop for the film at a temperature $T^* = 1.0$ subject to an applied oscillatory field of amplitude $H_0 = 1.0$ at field sweep rate $FSR = 1$, corresponding to a period of 240 MCSS for sinusoidal applied field. In addition Fig. 12 shows the hysteresis loops for the same system at lower angular frequencies of the applied oscillatory field with field sweep rates of $FSR = 10$, 100 and 1000. The first point to note is that the results show the same general trends previously seen in studies of the bulk kinetic Ising model [11,14]. The hysteresis loop area increases with decreasing angular frequency until it reaches a maximum before falling to zero. The dynamic phase transition occurs at a critical angular frequency $\omega_c$ close to where the hysteresis loop area is a maximum. For $\omega > \omega_c$, the system is in a dynamically ordered state. A dynamically disordered state is observed for $\omega < \omega_c$. Hysteresis loops that enclose the origin only occur for angular frequencies below $\omega_c$.

In Fig. 10, we show the results for $\omega < \omega_c$ and $T > T_c$. Starting from Fig. 8(c) where $FSR = 1$, the loop area decreases with increasing $FSR$ i.e. decreasing angular frequency of the applied oscillatory field. At the lowest angular frequency studied with $FSR = 1000$ (Fig. 10(c)), the hysteresis loop area is substantially smaller than for $FSR = 1$ (Fig. 8(c)), an angular frequency 1000 times smaller. However note that the qualitative shape of the hysteresis loop changes for $FSR = 1$ where there are no "tails" to the hysteresis curve that correspond to saturation of the magnetization. This indicates that the angular frequency $\omega = 2\pi / 240$ is close to the critical frequency $\omega_c$ for the film.

For fixed temperature and amplitude of the applied oscillatory field, the hysteresis loop area shows a power dependence on the angular frequency of the applied sinusoidal field with $\langle A \rangle \sim \omega^\beta$. Over the range $1 < FSR < 1000$, for $T^* = 1.0$ and $H_0 = 1.0$ the results in this paper give an exponent $\beta = 0.47$ which is comparable with the exponent obtained for the three dimensional Ising model [11]. Although, it should be noted that any estimate for the power law-scaling exponent $\beta$ determined from a restricted frequency range must be treated with caution [14].

## VII. Conclusion

The dynamic response of thin ferromagnetic Heisenberg films with competing surface fields to an applied oscillatory field has been studied. The magnetic spins in the model are continuously orientable, but the bilinear exchange anisotropy $\Lambda$ in the Heisenberg Hamiltonian ensures that Ising-like characteristics are retained. The competition between the ferromagnetic ordering tendencies of the spins and the applied oscillatory field determines the behavior of the film, which exhibits a dynamic phase transition between dynamically ordered and dynamically disordered phases. The critical temperature of the dynamic phase transition, $T_{cd}$, is a function of the angular frequency $\omega$ and amplitude $H_0$ of the applied oscillatory field. Hysteresis loops centered on the origin are observed at temperatures above $T_{cd}$. But for $T < T_{cd}$ the film is in a dynamically ordered state and the hysteresis loop is displaced from the origin being located in either the positive or negative magnetization half plane depending on the initial conditions of the simulation. A study of the time-dependent layer magnetization across the film $M_n^z(t)$, has shown that for $T < T_{cd}$ the dynamically ordered and dynamically disordered phases coexist within the film. In the presence of competing surface fields, the critical temperature of the ferromagnetic – paramagnetic transition of the film is suppressed from the bulk system value to the interface localization – delocalization temperature $T_{ci}$. This work shows that $T_{cd} < T_{ci}$.

**Acknowledgments**

The authors would like to thank Carol K. Hall for her support and interest in this work. Comments by Per Arne Rikvold and Bikas K. Chakrabarti on an early version of the manuscript are gratefully acknowledged.

## References

[1] Y.-L. He and G.-C. Wang, Phys. Rev. Lett. **70**, 2336 (1993).
[2] Q. Jiang, H.-N. Yang, and G.-C. Wang, Phys. Rev. B **52**, 14911 (1995).
[3] Jih-Shin Suen and J. L. Erskine, Phys. Rev. Lett. **78**, 3567 (1997).
[4] Jih-Shin Suen, M. H. Lee, G. Teeter, and J. L. Erskine, Phys. Rev. B **59**, 4249 (1999).
[5] T. Tomé and M. J. de Oliveira, Phys. Rev. A **41**, 4251 (1990).
[6] C. N. Luse and A. Zangwill, Phys. Rev. E **50**, 224 (1994).
[7] W. S. Lo and R. A. Pelcovits, Phys.Rev. A **42**, 7471 (1990).
[8] M. Acharyya and B. K. Chakrabarti, Phys.Rev. B **52**, 6550 (1995).
[9] M. Acharyya, Phys. Rev. E **56**, 1234 (1997).
[10] M. Acharyya, Phys. Rev. E **58**, 179 (1998).
[11] B. K. Chakrabarti and M. Acharyya, Rev. Mod. Phy. **71**, 847 (1999).
[12] S. W. Sides, P. A. Rikvold, and M. A. Novotny, Phys. Rev. E **57**, 6512 (1998).
[13] S. W. Sides, P. A. Rikvold, and M. A. Novotny, Phys. Rev. Lett. **81**, 834 (1998).
[14] S. W. Sides, P. A. Rikvold, and M. A. Novotny, Phys. Rev. E **59**, 2710 (1999).
[15] G. Korniss, C. J. White, P. A. Rikvold, and M. A. Novotny, Phys. Rev. E **63**, 016120 (2001).
[16] H. Fujisaka, H. Tutu, and P. A. Rikvold, Phys.Rev. E **63** 036109 (2001).
[17] K. Binder, D. P. Landau, and A. M. Ferrenberg, Phys. Rev. Lett. **74**, 298 (1995).
[18] K. Binder, D. P. Landau, and A. M. Ferrenberg, Phys. Rev. E **51**, 2823 (1995).
[19] K. Binder, R. Evans, D. P. Landau, and A. M. Ferrenberg, Phys. Rev. E **53**, 5023 (1996).
[20] A. M. Ferrenberg, D. P. Landau and K. Binder, Phys. Rev. E **58**, 3353 (1998).
[21] H. Jang and M. J. Grimson, Phys. Rev. B **55**, 12556 (1997).
[22] H. Jang and M. J. Grimson, J. Phys: Condens. Matter **10**, 9641 (1998).
[23] K. Binder and D. P. Landau, Phys. Rev. B **13**, 1140 (1976).
[24] N. Metropolis, A.W. Rosenbluth, M. N. Rosenbluth, A. H. Teller, and E. Teller, J. Chem. Phys. **21**, 1087 (1953).
[25] J. R. Barker and R. O. Watts, Chem. Phys. Lett. **3**, 144 (1969).
[26] L. C. Sampio, M. P. de Albuquerque, and F. S. de Menezes, Phys. Rev. B **54**, 6465 (1996).
[27] P. A. Rikvold, H. Tomita, S. Miyashita, and S. W. Sides, Phys. Rev. E **49**, 5080 (1994).

**Figure Captions**

**Fig. 1** Period-averaged magnetization, $\langle Q \rangle$, (solid symbols) and hysteresis loop area, $\langle A \rangle$, (open symbols) as a function of the amplitude of applied oscillatory field, $H_0$, at a temperature of $T^* = 1.0$ in the bulk system (circle) and the free film with no surface fields (triangle).

**Fig. 2** Period-averaged magnetization, $\langle Q \rangle$, (solid symbols) and hysteresis loop area, $\langle A \rangle$, (open symbols) as a function of the temperature $T^*$ for an applied oscillatory field amplitude of $H_0 = 1.0$ in the bulk system (circle) and the free film with no surface fields (triangle).

**Fig. 3** Period-averaged magnetization, $\langle Q \rangle$, (solid symbols) and hysteresis loop area, $\langle A \rangle$, (open symbols) as a function of the amplitude of applied oscillatory field, $H_0$, for temperatures of $T^* = 0.6$ (circle), $T^* = 1.0$ (triangle) and $T^* = 1.2$ (square).

**Fig. 4** Period-averaged magnetization, $\langle Q \rangle$, (solid symbols) and hysteresis loop area, $\langle A \rangle$, (open symbols) as a function of the temperature $T^*$ for applied oscillatory field amplitudes of $H_0 = 0.3$ (circle), $H_0 = 0.55$ (triangle) and $H_0 = 1.0$ (square).

**Fig. 5** Dynamic response of $z$ component of the magnetization, $M_z(t)$, at a temperature of $T^* = 1.0$ for applied oscillatory field amplitudes of $H_0 = 0.3, 0.7, 1.0, 2.0, 3.0$.

**Fig. 6** Hysteresis loop at a temperature of $T^* = 1.0$ for applied oscillatory field amplitudes of (a) $H_0 = 0.3$, (b) $H_0 = 0.7$, (c) $H_0 = 2.0$, and (d) $H_0 = 3.0$.

**Fig. 7** Dynamic response of the $z$ component of the magnetization, $M_z(t)$, to an applied oscillatory field of amplitude $H_0 = 1.0$ at temperatures of $T^* = 0.6, 0.8, 1.0, 1.2, 1.4$.

**Fig. 8** Hysteresis loop for an applied oscillatory field of amplitude $H_0 = 1.0$ at temperatures of (a) $T^* = 0.6$, (b) $T^* = 0.8$, (c) $T^* = 1.0$, and (d) $T^* = 1.4$.

**Fig. 9** Dynamic response of the layer magnetization across the film, $M_n^z$, for (a) $H_0 = 1.0$ and $T^* = 1.0$, (b) $H_0 = 1.0$ and $T^* = 0.6$, and (c) $H_0 = 3.0$ and $T^* = 1.0$.

**Fig. 10** Hysteresis loop for an applied oscillatory field of amplitude $H_0 = 1.0$ at a temperature of $T^* = 1.0$ for field sweep rates of (a) 10, (b) 100, and (c) 1000.

Fig. 1

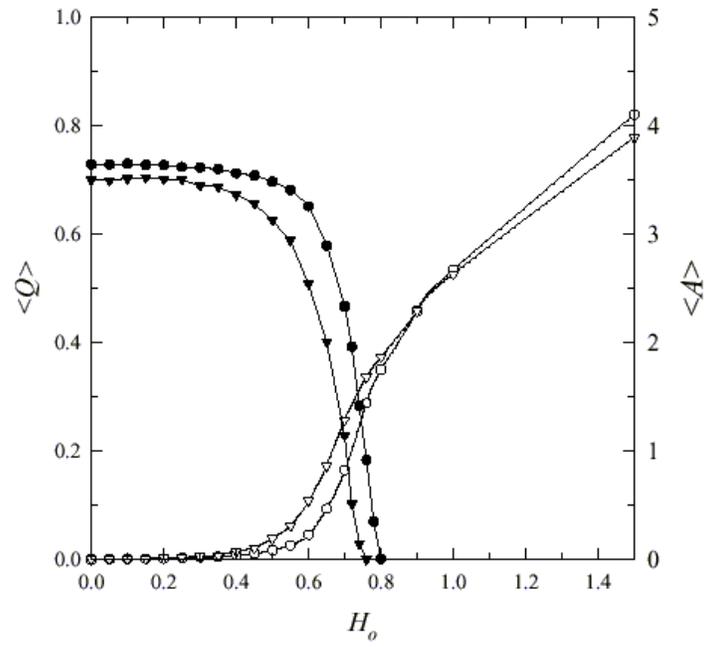

Fig. 2

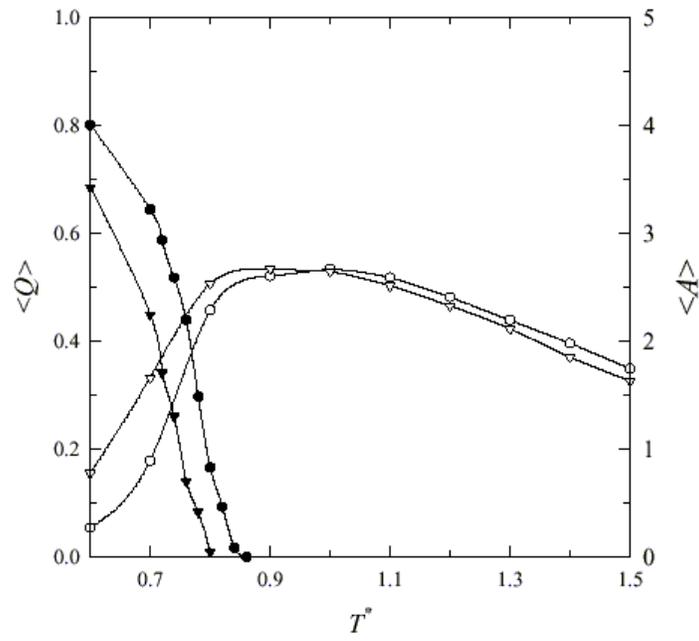

Fig. 3

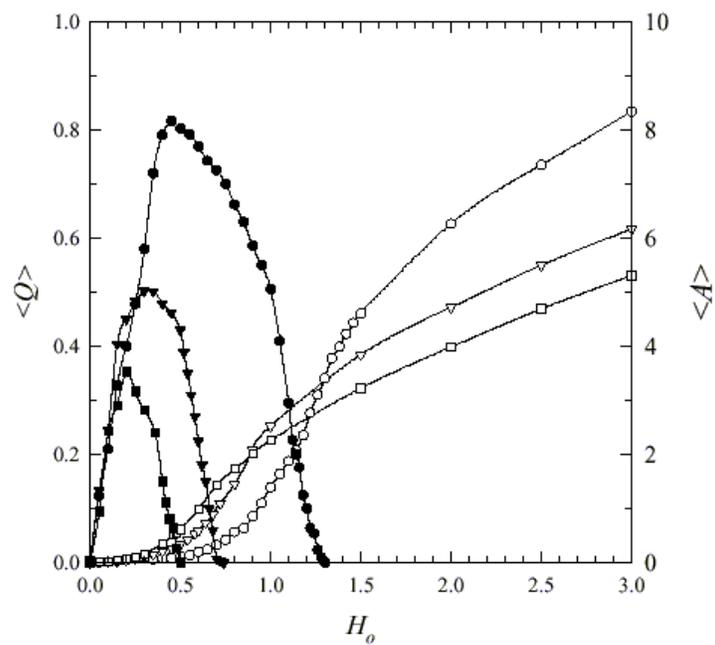

Fig. 4

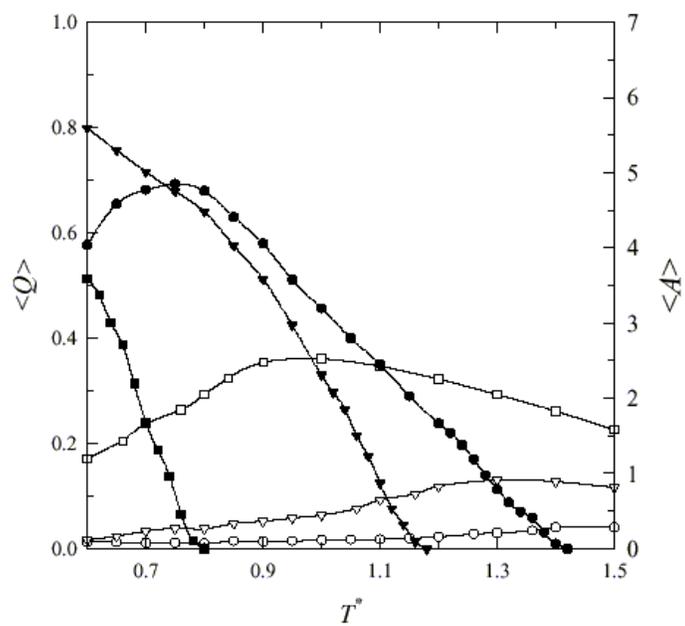

Fig. 5

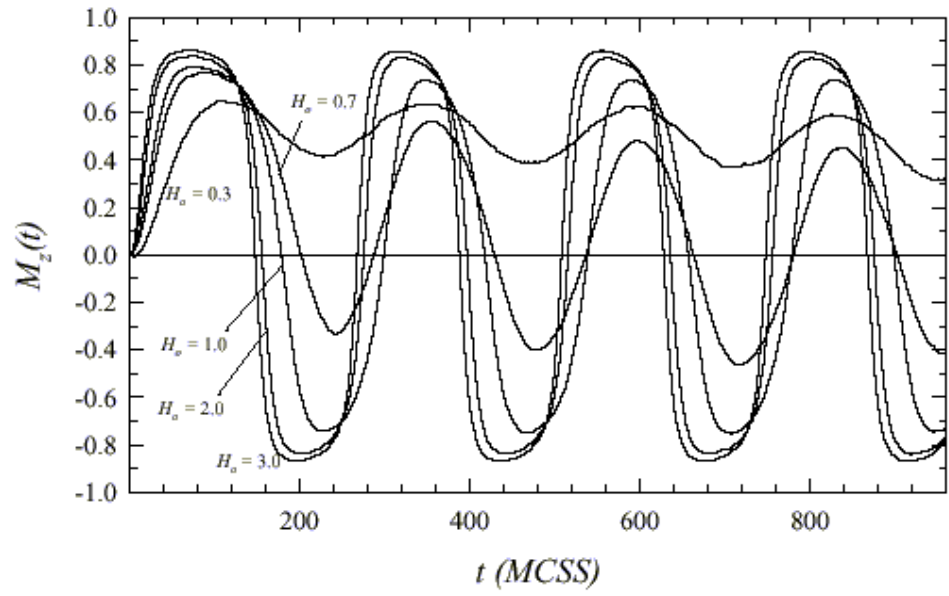

Fig. 6(a)

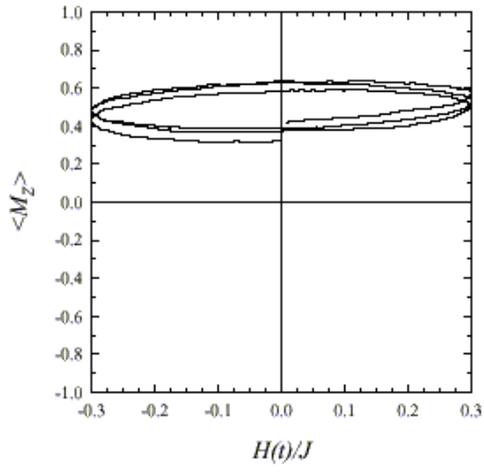

Fig. 6(c)

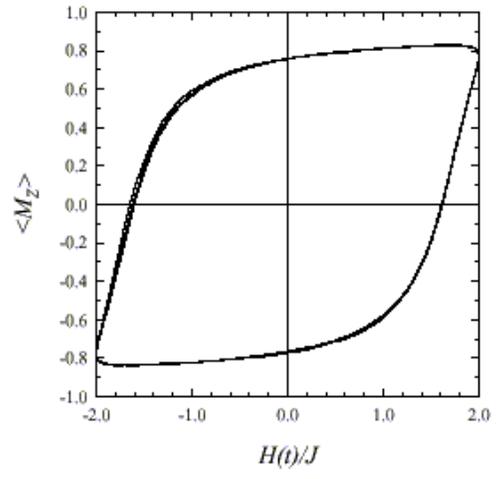

Fig. 6(b)

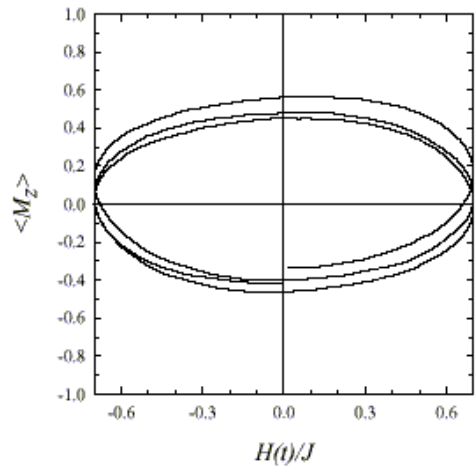

Fig. 6(d)

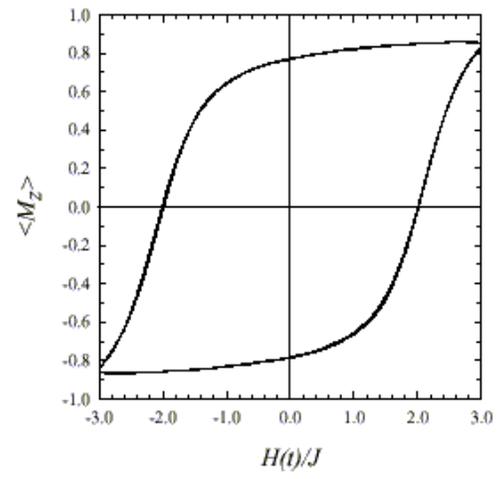

Fig. 7

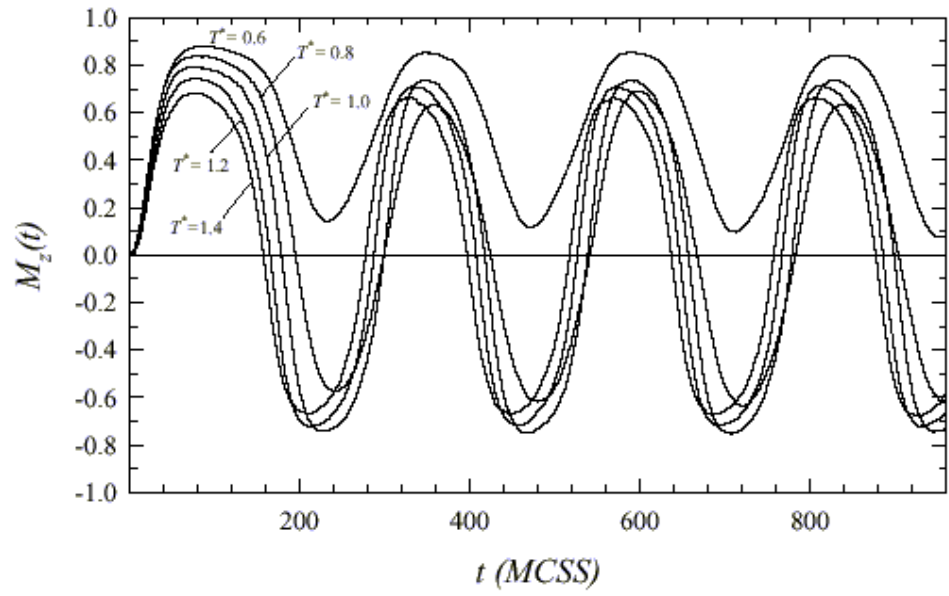

Fig. 8(a)

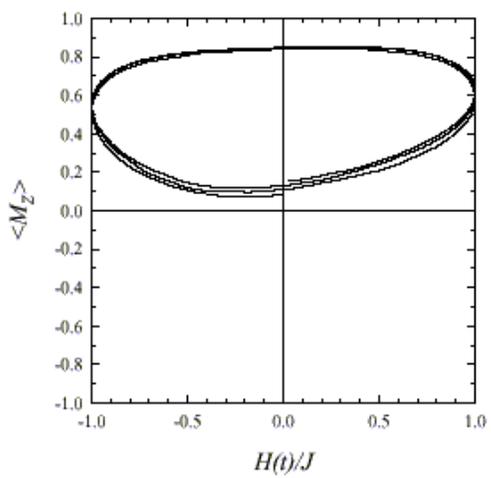

Fig. 8(b)

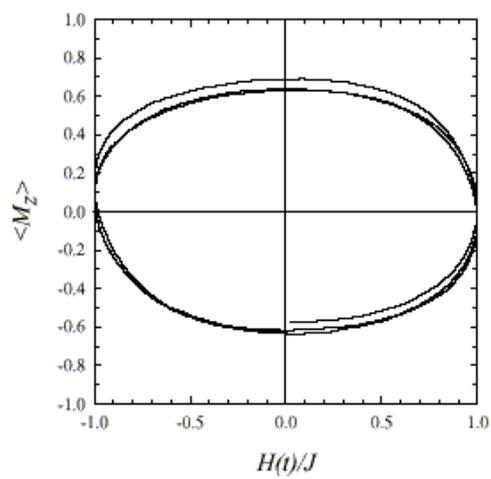

Fig. 8(c)

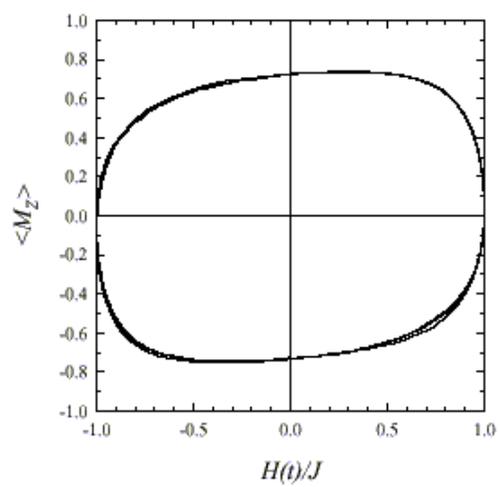

Fig. 8(d)

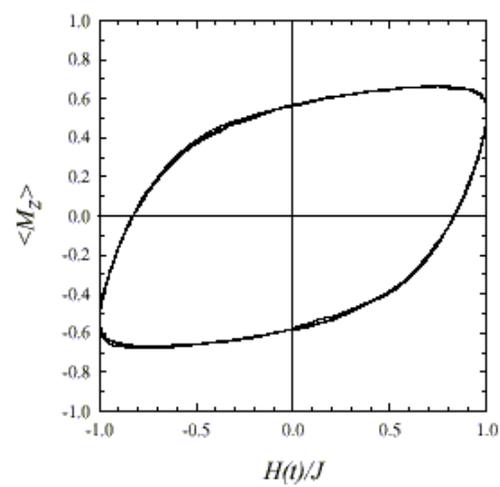

Fig. 9(a)

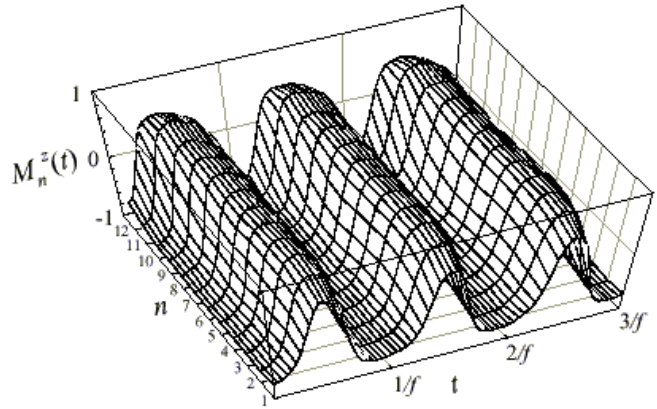

Fig. 9(b)

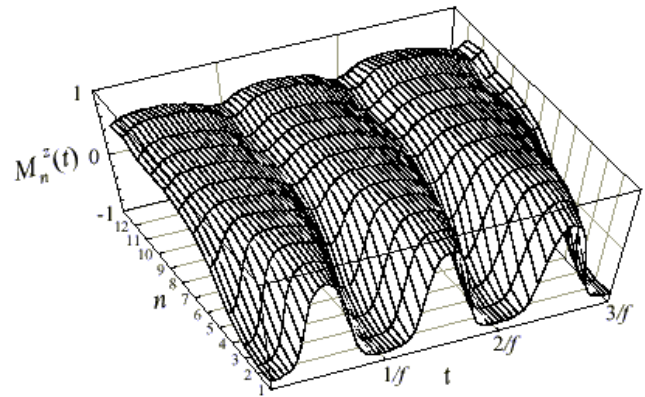

Fig. 9(c)

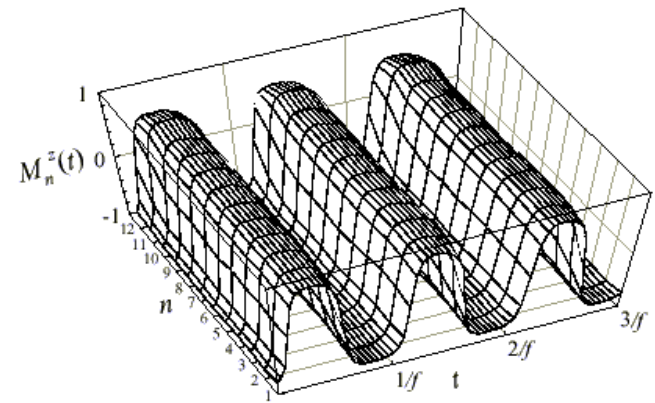

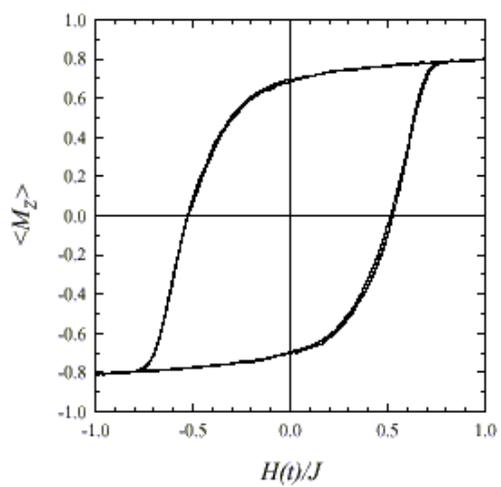

Fig. 10(a)

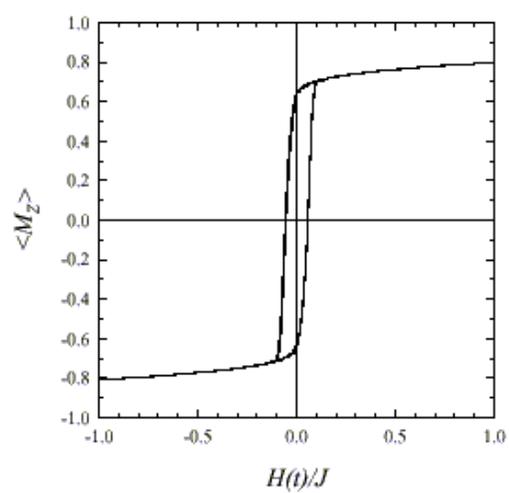

Fig. 10(c)

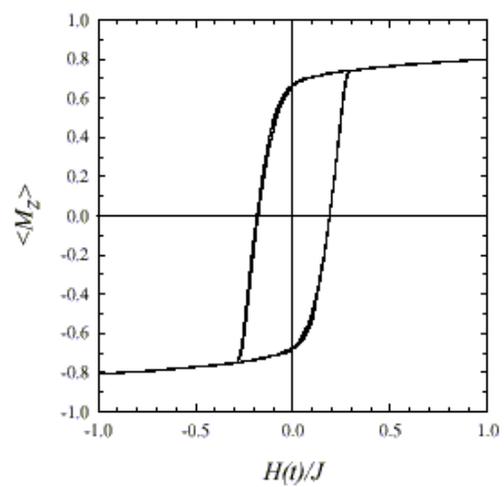

Fig. 10(b)